\documentclass[twocolumn,showpacs,prl,aps]{revtex4}
\usepackage{amssymb}
\usepackage{amsmath}
\usepackage{amsfonts}
\usepackage{graphics}
\usepackage{epic}
\usepackage{eepic}
\usepackage{color}
\usepackage{epsfig}
\usepackage{bbm}
\usepackage{graphicx}

\begin{document}

\def\ra{\rangle}
\def\la{\langle}
\def\bege{\begin{equation}
  }

\def\ende{\end{equation}}

\def\begarr{\begin{eqnarray}
  }

\def\endarr{\end{eqnarray}}
\def\ha{{\hat a}}
\def\hb{{\hat b}}
\def\hu{{\hat u}}
\def\hv{{\hat v}}
\def\hc{{\hat c}}
\def\hd{{\hat d}}
\def\no{\noindent}\def\non{\nonumber}
\def\hi{\hangindent=45pt}
\def\v{\vskip 12pt}

\newcommand{\bra}[1]{\left\langle #1 \right\vert}
\newcommand{\ket}[1]{\left\vert #1 \right\rangle}

\title{Relating computational complexity and quantum spectral complexity}

\author{David R.\ Mitchell\footnote{E--mail address: david.mitchell@csun.edu}}
\affiliation{Department of Physics and Astronomy, California State University, Northridge, CA  91330, USA}
\date{\today}

\pacs{03.67.-a, 03.67.Lx, 05.45.Mt, 05.45.Pq }

\begin{abstract}
\noindent


It is found that the statistical level fluctuations of the AQC 3-SAT problem undergo a transition from a poisson (regular) fluctuation form to a form consistent with the predictions of Random Matrix Theory.  We present data which suggests this transition correlates with the computational phase transition in the classical 3-SAT problem.  Application to Gaussian Processes and implication for experiment is discussed.

\end{abstract}

\maketitle

\section{Introduction}

The statistical examination of level fluctuation properties in the spectra of quantum systems has a rich history beginning with the initiation of Random Matrix Theory (RMT) by Wigner \cite{Wigner1967} and developed by Dyson \cite{Dyson1965} and Mehta \cite{Mehta1991} to predict the statistical properties of nuclear spectra and later applied to spectra from atomic and condensed matter systems.  Such spectra are characterized notably by level repulsion and a strong correlation between energy levels and this earlier work took as essential the condition that one is dealing with a system containing a large number of degrees of freedom (many-particle systems).  Later, the domain of validity of RMT was enlarged to include the spectra of simple quantum systems possessing classical analogs characterized by irregular (chaotic) motion \cite{Bohigas198ba}\cite{Bohigas1991}.  Subsequent research led to the correlation between spectral properties predictable by RMT and the presence of positive Liapunov exponents in the corresponding classical dynamics in a large number of quantum systems possessing a chaotic classical analog.  Recently, the presence of level fluctuation properties consistent with RMT predictions was found in the spectra of Adiabatic Quantum Computing (AQC) algorithm for the 3-SAT boolean satisfiability problem, an NP-Complete problem \cite{Mitchelletal2005}.

In this letter, we examine the correlation between spectral complexity and classical computational algorithmic complexity in the AQC 3-SAT, a boolean satisfiability problem containing m clauses in $n$ variables, with each clause containing 3 boolean variables.  The classical computational complexity of this problem has been studied as a function of the clause-to-variable parameter\cite{Davisetal1962}\cite{Selmanetal1992}, and is characterized by a computational phase transition separating computationally `easy' and `hard' regions.  The spectral complexity is measured by the degree to which the corresponding adiabatic quantum computation spectrum takes on the spectral properties predicted by Random Matrix Theory.  In the following, we present initial results relating the classical and quantum measures of complexity for the 3-SAT problem.


\section{Background}

Central to the adiabatic quantum computational (AQC) algorithm \cite{Farhietal2000} is the adiabatic evolution of the ground state energy level, with this state evolving from a easily-constructable state to a final state that encodes the solution to a particular instance of the 3-SAT problem.  Until recently, little attention was paid to the characteristics of the entire spectrum.  However, such a characterization\cite{Mitchelletal2005} revealed the presence of level repulsion and fluctuation properties consistent with RMT in portions of the interpolation region for select problem parameters.


This presence of such spectral statistics enlarges further the domain of validity of RMT to include Quantum Computational systems.  It remains to  characterize the quantum problem via spectral regularity throughout problem parameter space.

Additionally, there has been extensive theoretical and experimental work examining systems that obye a Random Matrices type theory containing an additional adiabatic parameter dependence\cite{Altshuler}\cite{SimonsKleppAlt}\cite{Mitchelletal1996}.  Such systems, called Gaussian Processes, have a number of interesting properties including a universal scaling that brings two-point correlation functions of eigenvectors or eigenvalues from differing spectra to a common, universal form. Thus, the characterization of spectra as adhering to RMT would provide the first step in revealing possible properties associated with Gaussian processes in AQC.

This work parallels the correspondence between the global change in spectral complexity and classical dynamical order as measured by the Liapunov dynamical exponent central to the study of quantal chaos in the 1980s and 1990s \cite{Bohigas1991}\cite{Bohigas198ba}, extending ideas regarding the Bohr Correspondence Principle as applied to classically chaotic systems.


This paper is organized as follows:  we begin with an examination of Random Matrix Theory and spectral measures followed by a discussion of the quantum adiabatic computational algorithm.  This is followed by the results of the statistical analysis of direct diagonalization of the AQC 3-SAT spectrum for various problem parameters.  These results and their generalization to experiment are discussed in the context of the asymptotic large N (qubit) limit.


\section{Random Matrix Theory}

Random Matrix Theory (RMT) \cite{Mehta1991} has been useful in describing the statistical fluctuations of spectra in a variety of quantum systems.  Historically, RMT~\cite{Brodyetal1981} has found application describing statistical spectral characteristics of a variety of complex systems such as compound-nuclear resonances\cite{Wigner1967}, many-particle atomic spectra, weakly disordered systems \cite{Altshdisordered}, few-body atomic systems possessing a chaotic classical analog \cite{Bohigas198ba}\cite{klepp}, and even number theory in the distribution of the zeros of the Riemann zeta function\cite{Montgomery}\cite{Titchmarsh}\cite{Odlyzko}.
In addition, recent work\cite{Mitchelletal2005} has shown the additional applicability of RMT in selected instances from the AQC 3-SAT problem \cite{Mitchelletal2005}.

RMT is a statistical description of complex quantum systems in which detailed knowledge about particle interactions is abandoned in favor of a general ensemble description with the only restriction due to physical symmetries.  Systems conserving time reversal symmetry are described by the Gaussian orthogonal ensemble (GOE), while systems with broken time reversal symmetry are described by the Gaussian unitary ensemble (GUE).

When examining the global spectral character of a system, the average level density $\rho(E)$, a model dependent property, is normalized to unity.  When so normalized, the resulting {\em unfolded} spectra assumes universal properties, and the spectra from a large variety of physical systems may be compared, with theoretical predictions determined from the theory of Random Matrices.  In particular, the theoretical spectral nearest-neighbor spacing (NNS) fluctuation distribution assumes a universal form that may compared with the {\em statistical} properties of the experimental levels of interest.  Whether the specific distribution assumes universal properties depends on the presence of quantal symmetries in the system, and, in the case of quantal systems with classical analogs, RMT spectral predictions correlate with the presence of classical nonlinearities as distinguished, for example, by positive Liapunov exponents.

The NNS distribution typical for complex quantum and disordered systems is given by the Wigner distribution~\cite{Brodyetal1981,Mehta1991} as predicted by RMT.  Such distributions are typical for complex systems without quantal symmetries, which results in correlated, irregular energy spectra that exhibit level repulsion.  Conversely, Hamiltonians corresponding to classical systems subject to symmetries and conservation laws typically display uncorrelated energy spectra and  nearest-neighbor level spacing distribution p(x) that is Poisson-like ($p(x) \sim exp(-x)$).

The spectral regularity may be quantified by the Brody parameter $q$ \cite{Brody1973}, occurring in a one-parameter distribution that interpolates between a regular Poisson spectrum $(q=0)$ and an irregular Wigner distribution $(q=1)$~\cite{Brody1973}. A renormalized spectrum having Brody parameter $q$  is characterized by the following NNS probability distribution for level spacing $\delta$, for the GOE:
\begin{equation}
p_q(\delta)=(1+q)\beta\delta^q\exp({-\beta\delta^{1+q})}\;,\;
\beta=\left[\Gamma\left(\frac{2+q}{1+q}\right)\right]^{1+q}\;.\label{fit}
\end{equation}

The AQC formulation is defined in terms of an adiabatic parameter that interpolates between an initial and final Hamiltonian.  In the theoretical limit of interpolation between two independent random matrices, parametric random matrix theory, or Gaussian processes (GP) is invoked.  Thus, one parameter disordered spectra is compared with the theoretical predictions of Gaussian processes (GP) whose two point correlation function assume a universal form when the parameter $x$ is rescaled by the average spectral quantity $\epsilon_i$, $x\rightarrow \sqrt{\overline{\left ( \frac{\partial \epsilon_i}{\partial x} \right)^2}} x$  as shown in the context of parametric RMT (GP) \cite{Mitchelletal1996} and derived also in the context of supersymmetry models \cite{Altshuler}.  Such spectra are characterized by an abundance of avoided level crossings, such as the spectra shown for the present case of the AQC 3-SAT problem shown in Fig. \ref{spectrum}.  Additional universal results have been obtained for multiparameter GP that relates to the occurrence of Berry phase in such systems and the (constant) universal density of Berry points in such systems when rescaled as above.


The GP in one parameter can be defined in a manner similar to RMT through its first two moments\cite{Mitchelletal1996},
\begin{eqnarray}\label{gproc}
\overline{H_{ij}(x)} & = & 0, \nonumber \\
\overline{H_{ij}(x) H_{kl}(x^\prime)} & = &
\frac{\omega^2}{2\beta}f(x-x^\prime)g^{(\beta)}_{ij,kl} \;
\end{eqnarray}
where $g^{(\beta=1)}_{ij,kl} =\delta_{ik}\delta_{jl}+\delta_{il}\delta_{jk}$, $g^{(\beta=2)}_{ij,kl}=2\delta_{il}\delta_{jk}$, $\omega$ is related to the Hamiltonian dimension and mean level spacing, and f(x) is normalized such that $f(0)=1$. $\beta =1$ is the GOE case, consisting of $H(x)$ that are real symmetric matrices, and $\beta = 2$ is the GUE case consisting of $H(x)$ that are complex Hermitian matrices.

A natural extension of RMT, Gaussian processes have been examined in the context of quantum dissipation \cite{Wilkinson1988}\cite{Wilkinson1990}\cite{Bulgacetal1996}, with a major component of the dissipation resulting from Landau-Zener nonadiabatic level transitions \cite{Zener1932} at avoided level crossings\cite{Schiff1955}.  This aspect is important to the adiabatic quantum computation algorithm, as algorithmic success requires the evolution of the system while remaining in the ground state, without such transitions.  As a result, interpolation through irregular spectral regions have a higher probability of transition and the interpolation must be slowed in this region when employing this approach to prevent level transitions.  From a computational standpoint, algorithmic cost increases with spectral complexity characterized by the onset of random matrix (Wigner) statistics.  For an AQC system possessing uniform average level spacing and obeying the conditions of a GP, it can be shown through the theory of GP that a quantum exponential speedup over classical algorithms is unlikely \cite{Mitchelletal2005}; however, such a result for the more realistic physical condition of a level density varying with energy is currently lacking.

\section{The Quantum Adiabatic Computation Algorithm}

The adiabatic algorithm \cite{Farhietal2000} allows for a quantum adiabatic process to perform a computation of an NP-Complete problem, the 3-SAT boolean satisfiability problem.  The time dependent Hamiltonian for such a process with an interpolation time $T$ takes the form
\begin{equation} \label{Farhi}
H(\frac{t}{T}) = (1-\frac{t}{T})H_b + \frac{t}{T}H_p ,
\end{equation}
where the Hamiltonian $H_b$ has the property that the ground state is easily constructed, and the Hamiltonian $H_p$ has the property that the ground state encodes the solution to the particular instance of the 3-SAT problem under investigation.

Starting with the ground state eigenstate of $H_b$, the approach allows for the calculation of a nontrivial computation problem through adiabatic interpolation of this state from the initial to final Hamiltonian.  Upon completion of the ground state interpolation, the state encodes the solution to the 3-SAT problem, which can be read through projective measurement of the state.  This approach has been shown to be equivalent to `traditional' circuit models of quantum computation\cite{vandam04}, in which quantum algorithmic speedup over classical approaches for a number of algorithms is well established.

The traditional measure of algorithmic complexity of the 3-SAT problem using classical computational measures involving either the Davis-Putnam \cite{Davisetal1962} or the GSAT algorithm\cite{Selmanetal1992}, shows a marked transition at a given clause-to-variable ratio, $f \thicksim 4.2$, indicating a computational phase transition from easy to hard solubility \cite{KirkpatrickSelman1994} in the large $n$ limit.  For problem sizes having $n \lesssim 20$, problem sizes typical in numerical simulation, the transition phenomenon still exists, though the transition point is less well defined and somewhat displaced from its asymptotic value.

\section{Results}

\begin{figure}
\includegraphics[width=8cm]{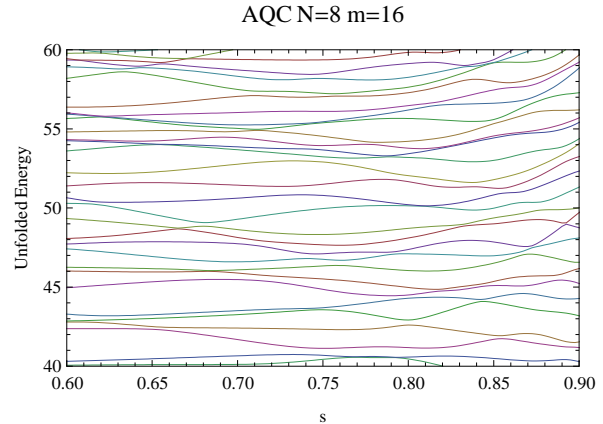}\\
\caption{Spectrum of irregular unfolded energies
of the instantaneous 3-SAT quantum computational Hamiltonian for N=8 m=16 clauses in the $N \sim [40,60]$ region. The predominance of avoided level crossings is characteristic of complex (irregular) spectra.}\label{spectrum}
\end{figure}

The statistical spectral fluctuation analysis involved determining the NNS distribution of the Hamiltonian $H(t)$, Eq. (\ref{Farhi}), at 100 points in the interpolation of a 3-SAT Hamiltonian with $n=8$ variables having matrix dimension $N=256$ for several values of the clause-to-variable ratio $f$.  In light of the above discussion relating computational cost and spectral complexity, we are interested in determining whether an irregular spectral region occurs in the interpolation and for each problem instance, the maximal Brody parameter is determined, with the results averaged over 200 problem instances for each $f$.

The results reveal a systematic change in the spectral regularity (Brody measure $q$) of the instantaneous AQC Hamiltonian H(t) during the course of the adiabatic quantum algorithm.  A typical unfolded (renormalized) spectrum associated with a particular problem instance is shown for a particular problem instance having $N=8$, $f=2$, shown in Fig. \ref{spectrum}.  In the initial phase of the interpolation for hard-to-solve problem instances, the statistical NNS fluctuations conform to a regular, Poisson-type distribution.  This corresponds to eigenvalue degeneracies inherent in the initial hamiltonian, $H(0)$.  Later in the interpolation, the fluctuation probability distribution transitions to an irregular, Wigner-type distribution.  It was found that such irregular spectra only occur for computational problem instances having ($f \gtrsim 1$).

\begin{figure}[t]
\includegraphics[width=8cm]{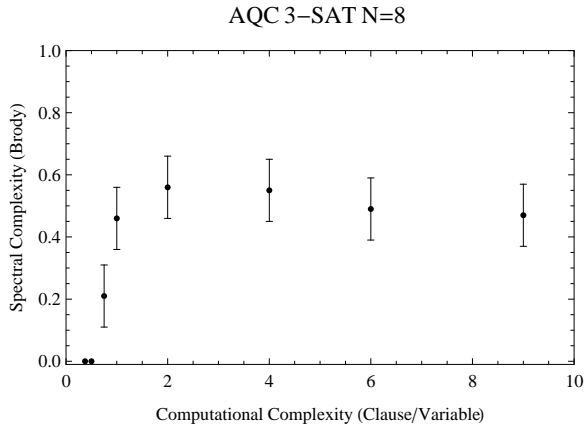}\\
\caption{Spectral Complexity, maximum Brody parameter, as a function of the classical computational complexity, clause-to-variable ratio, of the 3-SAT problem for N=8.}
\label{cxplot}
\end{figure}

Fig. \ref{cxplot} summarizes these quantum simulation results as a function of the classical algorithmic complexity parameter $f$, plotting spectral complexity (Brody measure) versus $f$ (clause-to-variable ratio).  For easy-to-solve computational problem instances ($f$ small), the spectral complexity was found to be zero.  For hard-to-solve instances of the 3-SAT problem the spectral complexity is nonzero.  The correlation between the ``classical'' computational complexity as parameterized by $f$ and the quantum spectral complexity, parameterized by the Brody parameter $q$ is shown in Fig. \ref{cxplot}.
We note that if the computational complexity is used as a proxy for computational cost, the quantum system appears to go through an orderly-complex transition in the region $1 \lesssim f \lesssim 4$ and diminishes somewhat for $f \gtrsim 5$.  The overall behavior exhibits the basic features of the classical algorithmic cost profile for the 3-SAT, which, for $n=8$, is characterized by a computational phase transition that is smoothed compared to the asymptotic form and displaced from the asymptotic (large $n$) phase transition point of 4.2.

Additional studies for $n$ closer to the asymptotic region ($n\gtrsim 20$) would be required to examine the asymptotic quantum transition and comparing the classical and quantum cost profiles in detail.  While the current results are for the $n=8$ non-asymptotic regime, the change in global spectral statistics correlated with computational complexity as measured by clause-to-variable ratio was found to be significant.  The salient feature is the transition to from regular to irregular spectra dependent on algorithmic complexity that correlates with classical results.

\section{Discussion}

Recent work \cite{Farhietal2000} showed that an NP-Complete problem can be converted into an adiabatic quantum mechanical process, which marked a new approach to quantum computation that was later shown equivalent to the circuit model of quantum computation\cite{vandam04}.

The present results enlarge the regime of validity of RMT to this additional class of AQC problems.  The central question we ask in this article is whether the spectral complexity is correlated with computational algorithmic complexity.

The results from Fig. \ref{cxplot} illustrate that when the entire spectrum from the AQC problem instances are examined, the spectrum is shown to be orderly for small clause-to-variable ratio, $f$, becoming more complex for larger $f$.  This leads us to the following result: when formulated in terms of an adiabatic quantum computation, all computationally hard NP-complete complexity class 3-SAT problems traverse a region characterized by an irregular spectrum.

How general is this result?  Our results verify the above conclusion for $n=8$ in regions that have been characterized as algorithmically complex ($f\sim4.2$), and we suggest that this result likely holds for larger $n$.  Spectra exhibiting orderly spectral statistics ($q \sim 0$) are typically associated with quantum symmetries, and it seems unlikely that a symmetry not manifesting itself for $n=8$ will manifest itself for larger $n$.

The above conjecture can be tested through a related experimental hypothesis related to the spectral statistics of the AQC problem:  there exists no classically hard instance of an NP-complete complexity class problem which, when formulated in terms of an adiabatic quantum computation, results in a simple spectrum (Poisson level spacing distribution) at all interpolation points in the computation.

Further, since quantum adiabatic computing has been shown equivalent to the circuit model of quantum computing (based on quantum bits) \cite{vandam04}, the above predictions would apply equally to all quantum computationally hard circuit model problems as well.

The experimental determination can be made by examining the spectrum at intermediate points in the interpolation for problem instances having fixed $n$ and varying clause-to-variable ratio.  This hypothesis would be {\it disproved} if a system were found that has the properties of (1) being NP-Complete, (2) computationally complex (hard-to-solve instance) determined by classical measures such as clause-to-variable ratio, and (3) the property of having a regular spectrum (Brody parameter $q=0$) at all interpolation points.  Such a system would correspond to an NP complete, hard-to-solve problem instance possessing a regular spectrum, which would require no interpolative slow down due to irregular (complex) spectral effects.

Finally, we have identified hard-to-solve instances of NP-Complete quantum adiabatic computation processes as having parameter regions obeying RMT and thus belonging to the class of Gaussian processes (GP).  Once this identification is made, a variety of universal results follow relating primarily to two-point correlation functions of wavefunctions and eigenvalues that have been tested in quantum systems possessing a classical analog and containing an external parameter dependence\cite{SimonsKleppAlt},\cite{KusMitch}.  If the above conjecture happens to be true, then spectral complexity would be correlated with computational algorithmic complexity and results consistent with GP should be found asymptotically in AQC processes having a problem parameter $f$ consistent with significan computational algorithmic complexity.

\vskip 0.5cm {\bf Acknowledgements} This research was initiated as an institute scholar at the Kavli Institute for Theoretical Physics (KITP), University of California, Santa Barbara, NSF Grant No. PHY99-07949.


\begin{thebibliography}{9}

\bibitem{Wigner1967}
E.P. Wigner, Random matrices in physics, SIAM Review {\bf 9}, 1-23 (1967).

\bibitem{Dyson1965}
F. J. Dyson, in {\it Statistical Theories of Spectra:
Fluctuations}, C. E Porter, ed., (Academic Press, New York, 1965).

\bibitem{Mehta1991}
M.L. Mehta, {\it Random Matrices}, (Academic Press, San Diego, 1991).

\bibitem{Bohigas1991}
O. Bohigas, Random matrices and chaotic dynamics, in {\it Chaos
and Quantum Physics}, M. Giannoni, A. Voros, and J. Zinn-Justin,
eds. (North-Holland, New York, 1991), pp. 87-199.; O. Bohigas, Chaotic motion and random matrix theories, in {\it Mathematical and Computational Methods in Nuclear Physics}, edited by J. S. Degesa et al. (Proceedings, Granada, Spain, 1983) Lecture Notes in Physics Vol. 209, pp. 1-99, Springer Verlag, Berlin (1984).

\bibitem{Bohigas198ba}
O. Bohigas, M. J. Giannoni, and C. Schmidt, Characterization of chaotic quantum spectra and universality of level fluctuation laws, {\it Phys. Rev. Lett.}{\bf 52} (1984) 1-4. O. Bohigas and M. J. Giannoni.

\bibitem{Mitchelletal2005} D. R. Mitchell, Christoph Adami, Waynn Lue, and Colin Williams, Random matrix model of quantum computing, Phys. Rev. A {\bf 71} (2005) 052324.

\bibitem{Davisetal1962}
M. Davis, G. Logemann, and D. Loveland, Comm. ACM {\bf 5}, 394 (1962).

\bibitem{Selmanetal1992}
B. Selman, H. J. Levesque, and D. G. Mitchell, A new method for solving hard satisfiability problems, in {\it Proceedings of the Tenth National Conference on Artificial Intelligence (AAAI-92)},
440-446. (AAAI Press/MIT Press, 1992).

\bibitem{Farhietal2000}
E. Farhi, J. Goldstone, S. Gutmann, J. Lapan, A. Lundgren, and D.
Preda. A quantum adiabatic evolution algorithm applied to random instances of an NP-complete problem, Science {\bf 292}, 472-475 (2002).  E. Farhi, J. Goldstone, S. Gutmann, and M. Sipser, Quantum
computation by adiabatic evolution, quant-ph/0001106 (2000).

\bibitem{SimonsKleppAlt}
B.D. Simons, A. Hashimoto, M. Courtney, D. Kleppner, and B.L. Altshuler.  New class of universal correlations in the spectra of hydrogen in a magnetic field, Phs. Rev. Lett {\bf 71}, 2899-2902 (1993).

\bibitem{Altshuler}  B.D. Simons and B.L. Altshuler, Phys. Rev. Lett {\bf 70} (1993) 4063; {\bf 72} (1994) 64; Nucl. Phys. B {\bf 409} (1993) 487.

\bibitem{Mitchelletal1996}
D. Mitchell, Y. Alhassid, and D. Kusnezov, Gaussian processes and universal parametric decorrelations of wavefunctions, Phys. Lett. {\bf A 215}, 21 (1996).


\bibitem{Brodyetal1981}
T. A. Brody, J. Flores, J. B. French, P. A. Mello, A. Pandey, and
S. S. M. Wong, Random matrix physics: Spectrum and strength
fluctuations, Reviews of Modern Physics {\bf 53}, 385 (1981).

\bibitem{Altshdisordered} B.L. Altshuler and B.I. Shklovskii, Sov. Phys. JETP {\bf 91} (1986) 220.

\bibitem{klepp}
C. Iu, G.R. Welch, M.M. Kash, D. Kleppner, D. Delande, J.C. Gay, Physical Review Letters, 66, 145 (1991); C. Iu, G.R. Welch, M.M.Kash, L. Hsu, D. Kleppner, Physical Review Letters, (1989).

\bibitem{Montgomery} H. L. Montgomery, The Pair Correlation of zeros of the zeta function, in {it Proceedings of the Symposium on Pure Mathematics 24,} American Mathematical Society, Providence, R.I. (1973) pp. 181-193.

\bibitem{Titchmarsh} E. C. Titchmarsh, {\it The Theory of the Riemann Zeta Function,} Chapter 10, Clarendon Press, Oxford (1951).

\bibitem{Odlyzko} A. M. Odlyzko,  On the distribution of spacings between zeros of the zeta function, {\it Math. Comput.} {\bf 48} (1987) 273-308; The $10^20$-th zero of the Riemann zeta function and 70 million of its neighbors, AT\&T Bell Lab preprint (1989).

\bibitem{Brody1973}
T. A. Brody, Statistical measure for repulsion of energy levels, Lett. Nuovo Cimento {\bf 7}, 482 (1973).

\bibitem{KusMitch}  Dimitri Kusnezov and David Mitchell, Universal predictions for statistical nuclear correlations, Phys. Rev. C {\bf 54} (1996) 147; David Mitchell and Dimitri Kusnezov, Topological Dependence of Universal Correlations in Multiparameter Hamiltonians, Phys Rev E {\bf 54} (1996) 6207.

\bibitem{Wilkinson1988}
M. Wilkinson, Statistical aspects of dissipation by Landau-Zener
transitions, J. Phys. {\bf A 21}, 4021 (1988).

\bibitem{Wilkinson1990}
M. Wilkinson, Diffusion and dissipation in complex quantum
systems, Phys. Rev. {\bf A 41}, 4645 (1990).

\bibitem{Bulgacetal1996}
A. Bulgac, G. DoDang, and D. Kusnezov, Random matrix approach to quantum dissipation, Phys. Rev. {\bf E 54}, 3468 (1996).

\bibitem{Zener1932}
G. Zener, Non-adiabatic crossing of energy levels, Proc. Roy. Soc.
{\bf A 137}, 696 (1932).

\bibitem{Schiff1955}
L. Schiff, {\it Quantum Mechanics}, (McGraw-Hill, New York, 1955).

\bibitem{vandam04} D. Aharanov, W. van Dam, J. Kempe, Z. Landau, S. Lloyd, O. Regev, Adiabatic Quantum Computation is Equivalent to Standard Quantum Computation, SIAM Journal of Computing, Vol. 37, Issue 1, p. 166-194 (2007), conference version in Proc. 45th FOCS, p. 42-51 (2004)

\bibitem{KirkpatrickSelman1994}
S. Kirkpatrick and B. Selman, Critical behavior in the
satisfiability of random Boolean expresssions, Science {\bf 264}, 1297-1301 (1994).

\bibitem{Landauer} R. Landauer, Irreversibility and heat generation in the computing process, {\it IBM J. Res. Dev,} 5:183, 1961.

\end{thebibliography}
\end{document}